\documentclass[
  reprint,
  fleqn,
  superscriptaddress,
  preprintnumbers,
  amsmath,amssymb,
  aps,prl,
  floatfix
]{revtex4-2}
\usepackage[table,dvipsnames]{xcolor}
\definecolor{VividPurple}{HTML}{9B00FF} 
\definecolor{VividMagenta}{HTML}{FF00AA}
\definecolor{VividBlue}{HTML}{005CFF}
\definecolor{VividOrange}{HTML}{FF5500}
\definecolor{VividRed}{HTML}{FF0033}      
\definecolor{VividGreen}{HTML}{00CC22}    
\definecolor{VividCyan}{HTML}{00E5FF}     
\definecolor{VividYellow}{HTML}{FFEE00}   
\definecolor{VividTeal}{HTML}{009FFF}     
\definecolor{VividPink}{HTML}{FF0088}     
\definecolor{VividIndigo}{HTML}{4B00FF}   
\definecolor{VividLime}{HTML}{A4FF00}     
\definecolor{Emerald}{HTML}{50C878}
\usepackage{placeins}
\usepackage{amsmath}
\usepackage[british]{babel}
\usepackage{graphicx}
\usepackage{dcolumn}
\usepackage{ragged2e}
\usepackage{bm}
\usepackage[version=4]{mhchem}
\usepackage{comment}
\usepackage{hepparticles}
\usepackage[utf8]{inputenc}
\usepackage{booktabs}
\usepackage{lipsum}
\usepackage{subcaption}
\usepackage{afterpage}
\usepackage{tikz}
\usepackage{hyperref}
\hypersetup{
    colorlinks=true,
    linkcolor= VividBlue,
    filecolor=cyan,      
    urlcolor= VividBlue,
    citecolor= VividBlue,
    pdftitle={59Cu paper},
    pdfpagemode=FullScreen,
    }
\usepackage{array}
\usepackage{soul}
\usepackage{orcidlink}

\begin{document}
\flushbottom
\raggedbottom


\title{\texorpdfstring{Direct Measurement of the $^{59}$Cu$(p,\alpha)^{56}$Ni Excitation Function to Constrain the NiCu Cycle Strength and Its Impact on Explosive Nucleosynthesis}{Direct Measurement of the 59Cu(p,alpha)56Ni Excitation Function to Constrain the NiCu Cycle Strength and Its Impact on Explosive Nucleosynthesis}}

\author{E. Lopez-Saavedra \orcidlink{0000-0001-9187-0404}}
\email{elopezsaavedra@anl.gov}
\affiliation{Physics\hspace{0.1cm}Division,\hspace{0.1cm}Argonne\hspace{0.1cm}National\hspace{0.1cm}Laboratory,\hspace{0.1cm}Lemont,\hspace{0.1cm}Illinois\hspace{0.1cm}60439,\hspace{0.1cm}USA}

\author{M. L. Avila \orcidlink{0009-0002-4051-9627}} 
\affiliation{Physics\hspace{0.1cm}Division,\hspace{0.1cm}Argonne\hspace{0.1cm}National\hspace{0.1cm}Laboratory,\hspace{0.1cm}Lemont,\hspace{0.1cm}Illinois\hspace{0.1cm}60439,\hspace{0.1cm}USA}

\author{W.-J. Ong \orcidlink{0000-0002-0945-8654}}
\affiliation{Nuclear\hspace{0.1cm}and\hspace{0.1cm}Chemical\hspace{0.1cm}Sciences\hspace{0.1cm}Division,\hspace{0.1cm}Lawrence\hspace{0.1cm}Livermore\hspace{0.1cm}National\hspace{0.1cm}Laboratory,\hspace{0.1cm}Livermore,\hspace{0.1cm}California\hspace{0.1cm}94550,\hspace{0.1cm}USA}

\author{P. Mohr \orcidlink{0000-0002-6695-9359}}
\affiliation{HUN-REN Institute for Nuclear Research (ATOMKI), H-4001 \hspace{0.1cm} Debrecen, Hungary}

\author{A. Psaltis \orcidlink{0000-0003-2197-0797}}
\affiliation{Department of Astronomy \& Physics, Saint Mary’s University, Halifax, NS B3H 3C3, Canada}

\author{S. Ahn \orcidlink{0000-0001-8190-4914}}
\affiliation{Center\hspace{0.1cm}for\hspace{0.1cm}Exotic\hspace{0.1cm}Nuclear\hspace{0.1cm}Studies,\hspace{0.1cm}Institute\hspace{0.1cm}for\hspace{0.1cm}Basic\hspace{0.1cm}Science,\hspace{0.1cm}Daejeon\hspace{0.1cm}34126,\hspace{0.1cm}Republic\hspace{0.1cm}of\hspace{0.1cm}Korea}

\author{H. Arora \orcidlink{0000-0001-6183-658X}}
\affiliation{Center\hspace{0.1cm}for\hspace{0.1cm}Exotic\hspace{0.1cm}Nuclear\hspace{0.1cm}Studies,\hspace{0.1cm}Institute\hspace{0.1cm}for\hspace{0.1cm}Basic\hspace{0.1cm}Science,\hspace{0.1cm}Daejeon\hspace{0.1cm}34126,\hspace{0.1cm}Republic\hspace{0.1cm}of\hspace{0.1cm}Korea}

\author{L. Balliet \orcidlink{0000-0002-1081-1259}}
\affiliation{Facility\hspace{0.1cm}for\hspace{0.1cm}Rare\hspace{0.1cm}Isotope\hspace{0.1cm}Beams\hspace{0.1cm}(FRIB),\hspace{0.1cm}Michigan\hspace{0.1cm}State\hspace{0.1cm}University,\hspace{0.1cm}East\hspace{0.1cm}Lansing,\hspace{0.1cm}MI\hspace{0.1cm}48824,\hspace{0.1cm}USA}

\author{K. Bhatt \orcidlink{0000-0003-0100-8736}}
\affiliation{University of Notre Dame, Dept.\hspace{0.1cm} of\hspace{0.1cm} Physics\hspace{0.1cm} and \hspace{0.1cm} Astronomy, Notre \hspace{0.1cm}Dame, IN \hspace{0.1cm}46556, \hspace{0.1cm}USA}

\author{S.M. Cha \orcidlink{0009-0000-5988-5956}}
\affiliation{Center\hspace{0.1cm}for\hspace{0.1cm}Exotic\hspace{0.1cm}Nuclear\hspace{0.1cm}Studies,\hspace{0.1cm}Institute\hspace{0.1cm}for\hspace{0.1cm}Basic\hspace{0.1cm}Science,\hspace{0.1cm}Daejeon\hspace{0.1cm}34126,\hspace{0.1cm}Republic\hspace{0.1cm}of\hspace{0.1cm}Korea}

\author{K. A. Chipps \orcidlink{0000-0003-3050-1298}}
\affiliation{Oak\hspace{0.1cm}Ridge\hspace{0.1cm}National\hspace{0.1cm}Laboratory,\hspace{0.1cm}Tennessee,\hspace{0.1cm}USA}

\author{J. Dopfer \orcidlink{0009-0007-9235-4824}}
\affiliation{Facility\hspace{0.1cm}for\hspace{0.1cm}Rare\hspace{0.1cm}Isotope\hspace{0.1cm}Beams\hspace{0.1cm}(FRIB),\hspace{0.1cm}Michigan\hspace{0.1cm}State\hspace{0.1cm}University,\hspace{0.1cm}East\hspace{0.1cm}Lansing,\hspace{0.1cm}MI\hspace{0.1cm}48824,\hspace{0.1cm}USA}

\author{I. A. Tolstukhin \orcidlink{0000-0002-6631-7479}}
\affiliation{Physics\hspace{0.1cm}Division,\hspace{0.1cm}Argonne\hspace{0.1cm}National\hspace{0.1cm}Laboratory,\hspace{0.1cm}Lemont,\hspace{0.1cm}Illinois\hspace{0.1cm}60439,\hspace{0.1cm}USA}

\author{R. Jain \orcidlink{0000-0001-9859-1512}}
\affiliation{Nuclear\hspace{0.1cm}and\hspace{0.1cm}Chemical\hspace{0.1cm}Sciences\hspace{0.1cm}Division,\hspace{0.1cm}Lawrence\hspace{0.1cm}Livermore\hspace{0.1cm}National\hspace{0.1cm}Laboratory,\hspace{0.1cm}Livermore,\hspace{0.1cm}California\hspace{0.1cm}94550,\hspace{0.1cm}USA}

\author{M.J. Kim \orcidlink{0000-0002-4372-5592}}
\affiliation{Center\hspace{0.1cm}for\hspace{0.1cm}Exotic\hspace{0.1cm}Nuclear\hspace{0.1cm}Studies,\hspace{0.1cm}Institute\hspace{0.1cm}for\hspace{0.1cm}Basic\hspace{0.1cm}Science,\hspace{0.1cm}Daejeon\hspace{0.1cm}34126,\hspace{0.1cm}Republic\hspace{0.1cm}of\hspace{0.1cm}Korea}
\affiliation{Extreme\hspace{0.1cm}Rare\hspace{0.1cm}Isotope\hspace{0.1cm}Science,\hspace{0.1cm}Institute\hspace{0.1cm}for\hspace{0.1cm}Rare\hspace{0.1cm}Isotope\hspace{0.1cm}Science\hspace{0.1cm}(IRIS),\hspace{0.1cm}1\hspace{0.1cm}Gukjegwahak-ro,\hspace{0.1cm}Yuseong-gu,\hspace{0.1cm}Daejeon\hspace{0.1cm}34000,\hspace{0.1cm}Republic\hspace{0.1cm}of\hspace{0.1cm}Korea}

\author{K. Kolos \orcidlink{0000-0002-1726-4171}}
\affiliation{Nuclear\hspace{0.1cm}and\hspace{0.1cm}Chemical\hspace{0.1cm}Sciences\hspace{0.1cm}Division,\hspace{0.1cm}Lawrence\hspace{0.1cm}Livermore\hspace{0.1cm}National\hspace{0.1cm}Laboratory,\hspace{0.1cm}Livermore,\hspace{0.1cm}California\hspace{0.1cm}94550,\hspace{0.1cm}USA}

\author{F. Montes \orcidlink{0000-0001-9849-5555}}
\affiliation{Facility\hspace{0.1cm}for\hspace{0.1cm}Rare\hspace{0.1cm}Isotope\hspace{0.1cm}Beams\hspace{0.1cm}(FRIB),\hspace{0.1cm}Michigan\hspace{0.1cm}State\hspace{0.1cm}University,\hspace{0.1cm}East\hspace{0.1cm}Lansing,\hspace{0.1cm}MI\hspace{0.1cm}48824,\hspace{0.1cm}USA}

\author{D. Neto \orcidlink{0000-0002-5397-7048}}
\affiliation{Physics\hspace{0.1cm}Division,\hspace{0.1cm}Argonne\hspace{0.1cm}National\hspace{0.1cm}Laboratory,\hspace{0.1cm}Lemont,\hspace{0.1cm}Illinois\hspace{0.1cm}60439,\hspace{0.1cm}USA}
\affiliation{Department\hspace{0.1cm}of\hspace{0.1cm}Physics,\hspace{0.1cm}University\hspace{0.1cm}of\hspace{0.1cm}Illinois\hspace{0.1cm}Chicago,\hspace{0.1cm}845\hspace{0.1cm}W.\hspace{0.1cm}Taylor\hspace{0.1cm}St.,\hspace{0.1cm}Chicago,\hspace{0.1cm}IL\hspace{0.1cm}60607,\hspace{0.1cm}USA}

\author{S. D. Pain \orcidlink{0000-0003-3081-688X}}
\affiliation{Oak\hspace{0.1cm}Ridge\hspace{0.1cm}National\hspace{0.1cm}Laboratory,\hspace{0.1cm}Tennessee,\hspace{0.1cm}USA}

\author{J. Pereira \orcidlink{0000-0002-3934-0876}}
\affiliation{Facility\hspace{0.1cm}for\hspace{0.1cm}Rare\hspace{0.1cm}Isotope\hspace{0.1cm}Beams\hspace{0.1cm}(FRIB),\hspace{0.1cm}Michigan\hspace{0.1cm}State\hspace{0.1cm}University,\hspace{0.1cm}East\hspace{0.1cm}Lansing,\hspace{0.1cm}MI\hspace{0.1cm}48824,\hspace{0.1cm}USA}

\author{J. S. Randhawa \orcidlink{0000-0001-6860-3754}}
\affiliation{Department of Physics and Astronomy, Mississippi\hspace{0.1cm}State\hspace{0.1cm}University,\hspace{0.1cm}Mississippi\hspace{0.1cm}State,\hspace{0.1cm}MS\hspace{0.1cm}39762, USA}

\author{L. J. Sun \orcidlink{0000-0002-1619-7448}}
\affiliation{Facility\hspace{0.1cm}for\hspace{0.1cm}Rare\hspace{0.1cm}Isotope\hspace{0.1cm}Beams\hspace{0.1cm}(FRIB),\hspace{0.1cm}Michigan\hspace{0.1cm}State\hspace{0.1cm}University,\hspace{0.1cm}East\hspace{0.1cm}Lansing,\hspace{0.1cm}MI\hspace{0.1cm}48824,\hspace{0.1cm}USA}

\author{C. Ugalde \orcidlink{0000-0003-2721-6728}}
\affiliation{Department\hspace{0.1cm}of\hspace{0.1cm}Physics,\hspace{0.1cm}University\hspace{0.1cm}of\hspace{0.1cm}Illinois\hspace{0.1cm}Chicago,\hspace{0.1cm}845\hspace{0.1cm}W.\hspace{0.1cm}Taylor\hspace{0.1cm}St.,\hspace{0.1cm}Chicago,\hspace{0.1cm}IL\hspace{0.1cm}60607,\hspace{0.1cm}USA}

\author{L. Wagner \orcidlink{0009-0000-9954-9658}}
\affiliation{Facility\hspace{0.1cm}for\hspace{0.1cm}Rare\hspace{0.1cm}Isotope\hspace{0.1cm}Beams\hspace{0.1cm}(FRIB),\hspace{0.1cm}Michigan\hspace{0.1cm}State\hspace{0.1cm}University,\hspace{0.1cm}East\hspace{0.1cm}Lansing,\hspace{0.1cm}MI\hspace{0.1cm}48824,\hspace{0.1cm}USA}

\date{\today}

\begin{abstract}
A new direct measurement of the $^{59}\mathrm{Cu}(p,\alpha){}^{56}\mathrm{Ni}$ excitation function from 2.43--5.88 MeV in the center-of-mass was performed in inverse kinematics using the high-efficiency MUSIC active-target detector at FRIB. 
This reaction plays a critical role in constraining the strength of the NiCu cycle in different explosive astrophysical scenarios such as Type I X-ray bursts and the $\nu$p-process in neutrino-driven winds after a core-collapse supernova. The newly derived stellar rate is systematically lower than previous estimates, reducing NiCu-cycle recycling in X-ray bursts to below $0.74\%$ and enhancing the $\nu p$-process efficiency across the relevant temperature range, extending its operation up to $T_9 = 3.94^{+0.99}_{-0.85}$ while constraining the predicted endpoint.

\end{abstract}

\maketitle

Explosive nucleosynthesis in proton-rich environments occurs in several 
astrophysical scenarios, including Type~I X-ray bursts (XRBs) on 
accreting neutron stars, where the $rp$-process~\cite{WallaceWoosley1981,Woosley1976} 
operates, and the proton-rich, neutrino-driven winds following 
core-collapse supernovae (CCSNe), where the $\nu p$-process takes 
place~\cite{PARIKH2013225,Frolich_nup,Arcones2012NuP}.
Understanding these astrophysical environments requires a detailed knowledge of nuclear physics, including reaction rates, nuclear structure, and decay properties~\cite{THIELEMANN200774, Thielemann2010Processes,annurevKap,Langanke2004NuclearAstrophysics}.
In both scenarios, the production of proton-rich nuclei proceeds through extended sequences of proton captures on unstable isotopes, but while the $rp$-process is dominated by $(p,\gamma)$ reactions and subsequent $\beta^{+}$ decays, the $\nu p$-process relies on a combination of $(p,\gamma)$ and $(n,p)$ reactions. \par

The $\nu p$-process was first proposed by Fr\"ohlich \textit{et 
al.}~\cite{Frolich_nup} as a primary process that could explain the 
large Sr/Fe ratios observed in hyper-metal-poor stars~\cite{Frebel2005} 
and the enhanced Sr, Y, and Zr abundances inferred from galactic 
chemical evolution~\cite{Travaglio2004}; it is also a leading candidate 
to account for the solar abundances of light $p$-nuclei such as 
$^{92,94}$Mo and $^{96,98}$Ru, which are underproduced by the 
$\gamma$-process in CCSNe~\cite{Nishimura2019}. The $\nu p$-process operates 
in the proton-rich ($Y_e > 0.5$, where $Y_e$ is the electron fraction, 
defined as the proton-to-baryon number ratio), neutrino-driven winds 
ejected from the proto-neutron star seconds after core collapse: it 
proceeds from the seed nucleus $^{56}$Ni, assembled in quasi-equilibrium 
at $T_9 \gtrsim 4$ (where $T_9$ is the temperature in units of 
$10^{9}$~K), and begins once quasi-equilibrium freezes out near 
$T_9 \approx 3$~\cite{Wanajo_2011}. Antineutrino captures on free protons then generate neutrons that, 
through $(n,p)$ reactions, bypass the long-lived $\beta^+$-decay 
waiting points at $^{56}$Ni and $^{64}$Ge, allowing the reaction flow 
to proceed beyond the Fe group~\cite{Frohlich2012_vpprocess,Arcones2012NuP}.\par

Whether nuclei beyond the Fe group are produced is controlled by the competition of reactions at $^{59}$Cu. The competition between $^{59}\mathrm{Cu}(p,\alpha){}^{56}\mathrm{Ni}$ and $^{59}\mathrm{Cu}(p,\gamma){}^{60}\mathrm{Zn}$ defines the NiCu cycle: a dominant $(p,\alpha)$ channel drives material back to the Ni region and confines the flow, whereas a dominant $(p,\gamma)$ channel enables breakout toward heavier nuclei~\cite{Kim_2022,Lam_2022}. As shown by Arcones \textit{et al.}~\cite{Arcones2012NuP}, the temperature where this transition occurs, typically around $T_9 \sim 3$ when REACLIB rates~\cite{Cyburt2010} are used, determines where breakout happens; a higher crossover temperature places it closer to the proto-neutron star, where stronger antineutrino flux boosts $(n,p)$ reactions and improves $\nu p$-process efficiency.\par

The same branching point is also important in XRBs, where the $rp$-process encounters the waiting points $^{56}$Ni and $^{60}$Zn~\cite{vanWormer1994}. Sensitivity studies identify this reaction pair as strongly affecting burst energy generation and ash composition in this mass region~\cite{Cyburt_2016,Meisel_2019,Sultana2025_XRB_sensitivity}. As the most common thermonuclear explosions in the Galaxy, XRBs are key probes of neutron-star structure and explosive nucleosynthesis~\cite{Woosley1976,TimingNeutronStars2021,PARIKH2013225,Degenaar2018,Parikh_2008,Steiner2010EoS}. 

Tighter constraints on the $^{59}\mathrm{Cu}(p,\alpha){}^{56}\mathrm{Ni}$ 
and $^{59}\mathrm{Cu}(p,\gamma){}^{60}\mathrm{Zn}$ reaction rates, and 
on their branching ratio, are therefore essential for reliably 
predicting breakout from the NiCu cycle in both the $rp$- and 
$\nu p$-processes.
The  first direct measurement of the $^{59}\mathrm{Cu}(p,\alpha)^{56}$Ni reaction was 
performed by Randhawa \textit{et al.}~\cite{Jaspreet59Cu}, who reported a single data point at $E_{\mathrm{cm}} = 6.0 \pm 0.3~\mathrm{MeV}$. Their measured cross section was found to  be a factor of $\sim$1.6--4 lower than statistical-model predictions, suggesting a suppressed $(p,\alpha)$ strength. 
 Two additional data points between 4 and 5 MeV have very recently become available~\cite{Bhathi25}, providing an experimentally constrained rate that indicates weak NiCu cycle recycling in XRBs, below 5$\%$ up to 1.5~GK. However, the inferred angle-integrated cross section is sensitive to the assumed angular distribution, leading to a model-dependent systematic in the derived rate.
Other attempts to constrain this reaction have relied on spectroscopic information of states in $^{60}\mathrm{Zn}$ \cite{Kim_2022,Lotay}. Ref.~\cite{Lotay} constrains the NiCu cycle contribution to the $rp$-process flow to a maximum of 38\%, whereas Ref.~\cite{Kim_2022} estimates a much lower cycling of $\le 0.01\%$. However, the resulting constraints suffer from significant uncertainties, primarily due to the lack of experimental data on the corresponding $\alpha$-particle partial widths. Further experiments to study the NiCu cycle are ongoing, reviewed recently in Ref.~\cite{Sun_2025}.

 In this work, a new direct measurement of the $^{59}\mathrm{Cu}(p,\alpha)^{56}\mathrm{Ni}$ excitation function, extending down to $E_{\rm cm} = 2.43$ MeV, is presented, providing the strongest constraints to date on the astrophysical reaction rate and the first direct experimental input at energies relevant to the $\nu p$-process. This significantly reduces the uncertainties for this key reaction in the NiCu cycle and enables a more accurate assessment of its impact on nucleosynthesis in  XRBs and the $\nu p$-process in CCSN.

The experiment was performed in inverse kinematics at the Facility for Rare Isotope Beams (FRIB) using the Multi-Sampling Ionization Chamber (MUSIC) detector~\cite{Carnelli2015MUSIC}. The 8.418 MeV/u $^{59}$Cu beam was obtained in-flight by fragmentation of the 240 MeV/u $^{64}$Zn primary beam. The $^{59}$Cu beam delivered to MUSIC had an average purity of approximately 94$\%$ and an intensity of $\sim9 \times 10^{3}$ particles per second (pps).
MUSIC was filled with methane gas at 440 Torr,  which completely stopped the beam inside of the detector. The only beam contaminant was $^{59}$Ni, contributing approximately $6\%$ of the total beam. 

The MUSIC detector is an active-target system with an anode segmented
into 18 strips, the middle 16 divided into left and right sections
\cite{Carnelli2015MUSIC}, enabling energy-loss measurement and particle
identification along the detector. When a $(p,\alpha)$ reaction occurs at
a given strip, the $^{56}$Ni recoil is distinguished from the unreacted
$^{59}\mathrm{Cu}$ beam by its lower energy loss; summing the energy
deposited in strips downstream of the reaction point enables a
$\Delta E$--$\Delta E$ separation of the reaction channels
\cite{LopezSaavedra_Supplemental}. 
 
Cross sections for eight strips corresponding to effective center-of-mass energies ($E_{\text{c.m.}}^{\text{eff}}$) between 2.43 and 5.88 MeV (Table~\ref{tab:xsec_results}) were extracted. These energies were derived from measured MUSIC beam-energy losses, corrected for thick-target yield effects~\cite{Cauldrons}.The uncertainty in the center-of-mass energy arises from the uncertainty in the beam energy (0.5\% FWHM) and the uncertainty in the beam’s energy loss in different strips of the detector (5\%) (see Ref.~\cite{LopezSaavedra_Supplemental} for details). 
The estimated total energy loss of the beam in each anode strip defines the center-of-mass energy binning, $\Delta E_{cm}$. Systematic uncertainties ($\Delta\sigma_{\mathrm{sys}}$, Table~\ref{tab:xsec_results}) were determined from the spread in extracted yields after varying event-selection criteria, including $\Delta E$-$\Delta E$ gates and energy-loss derivative thresholds.

\begin{table}[!htb]
\centering
\caption{Measured effective energies, energy bins, and cross sections with statistical and systematic uncertainties.}
\label{tab:xsec_results}
\begin{tabular}{c c c c c}
\toprule
$E_{\mathrm{cm}}^{eff}$\hspace{-0.07cm} (MeV)&$\Delta E_{cm}$\hspace{-0.07cm} (MeV)& $\sigma$\hspace{-0.07cm} (mb)&$\Delta\sigma_{\text{stat}}$\hspace{-0.07cm} (mb)& $\Delta\sigma_{\text{sys}}$\hspace{-0.07cm} (mb) \\
\midrule
5.88 (3) &$^{+0.18}_{-0.26}$ & $7.024$     & $0.220$   & $2.355$ \\
5.44 (3)& $^{+0.18}_{-0.27}$ & $3.412$     & $0.154$   & $0.612$ \\
4.98 (4)& $^{+0.19}_{-0.28}$ & $1.861$   & $0.113$   & $0.300$ \\
4.49 (4) & $^{+0.21}_{-0.28}$ & $0.619$     & $0.065$  & $0.092$ \\
4.02 (4) & $^{+0.19}_{-0.31}$ & $0.248$     & $0.042$  & $0.083$ \\
3.54 (4) & $^{+0.17}_{-0.37}$ & $0.082$ & $0.024$  & $0.014$ \\
2.98 (5) & $^{+0.19}_{-0.37}$ & $\llap{\tiny\textless}0.015$   & $-$ & $-$ \\
2.43 (5) & $^{+0.18}_{-0.42}$ & $\llap{\tiny\textless}0.011$   & $-$ & $-$ \\
\bottomrule
\end{tabular}
\end{table}

Statistical uncertainties for the first strips were determined assuming Poisson statistics. For very low-statistics strips, confidence intervals were estimated using the Feldman–Cousins approach~\cite{FeldmanCousins1998}. For Strips 8 and 9, the last two data points in Table~\ref{tab:xsec_results}, with 2 and 1 events, respectively, and an assumed background of 5 counts, have 95\% C.L. intervals for the events of $[0.00,\,2.49]$ and $[0.00,\,1.88]$.

\begin{figure}[!ht]
    \centering
    \includegraphics[width=\linewidth]{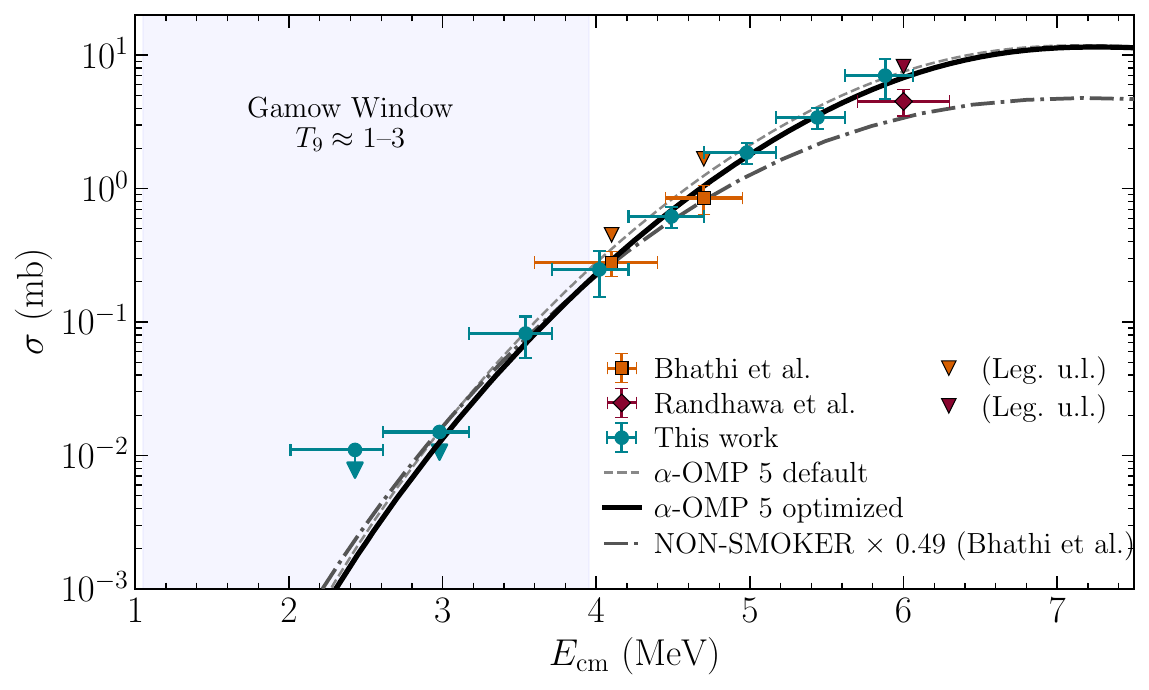}
   \caption{Measured $^{59}\mathrm{Cu}(p,\alpha)^{56}\mathrm{Ni}$ cross sections from this work (teal circles), Randhawa~\textit{et al.}~\cite{Jaspreet59Cu} (maroon diamond), and Bhathi~\textit{et al.}~\cite{Bhathi25} (orange squares: recommended values; triangles: Legendre upper limits), compared with the NON-SMOKER$\,\times\,0.49$ adopted by Bhathi \textit{et al.} (dark gray dash-dotted), and the DEM-3 ($\alpha$-OMP 5)~\cite{DEMETRIOU2002253} with optimized (black) and unmodified (gray dashed) geometry. Shaded region represents the Gamow window for temperatures $T_9\simeq1$--3}
    \label{fig:placeholder_xsec}
\end{figure}

Fig.~\ref{fig:placeholder_xsec} compares the measured cross sections with available experimental data~\cite{Bhathi25,Jaspreet59Cu}, the NON-SMOKER predictions scaled by a factor of 0.49 (adopted by Ref.~\cite{Bhathi25}), and TALYS~\cite{talys2} calculations using the Demetriou and Goriely 
dispersive $\alpha$-optical model potential ($\alpha$-OMP 5)~\cite{DEMETRIOU2002253}, shown both with its default geometry and with the optimized geometry  that best reproduces the data. The methodology used to identify this configuration is described briefly below and in detail in Ref.~\cite{LopezSaavedra_Supplemental}. Figure~\ref{fig:placeholder_xsec} also highlights the Gamow window for temperatures $T_9\simeq1$--3, which are relevant for the $\nu p$-process ($T_9\simeq1$--3) and XRBs ($T_9 \lesssim 1$).

The eight cross-section values reported here span $E_{\mathrm{c.m.}} = 2.43$--$5.88$~MeV and provide a detailed experimental mapping of the $^{59}$Cu$(p,\alpha)^{56}$Ni excitation function over this range, extending direct measurements to lower energies and reaching the high-temperature end of the $\nu p$-process window. At the energy closest to the Randhawa \textit{et al.}\ data point ($E_{\mathrm{c.m.}} \approx 6.0 \pm 0.3$~MeV), our measurement at $E_{\mathrm{c.m.}} = 5.88$~MeV is roughly a factor of 1.5 higher, but remains consistent within the quoted uncertainties. Ref.~\cite{Bhathi25} reported two alternative extractions of the angle-integrated cross section from the same dataset, based on TALYS-derived and Legendre-fit angular distributions; the resulting discrepancy illustrates a strong sensitivity to the angular-integration method. Our measurement, which directly yields the angle-integrated cross section, is free from this model dependence. Fig.~\ref{fig:placeholder_xsec} also shows that the NON-SMOKER prediction, even when scaled by the factor of 0.49 adopted in Ref.~\cite{Bhathi25}, does not reproduce the measured energy dependence, indicating that it is not an adequate description of the excitation function in this region.

The present measurement constrains the reaction rate over $T_9 \sim
2.6$--10.5, the effective temperature window determined numerically
following Ref.~\cite{Rauscher_PRC2010_gamow}. To extend the rate toward lower temperatures, the experimental (p,$\alpha$) cross sections were combined with TALYS
statistical-model calculations 

In a simplified notation following the Hauser-Feshbach statistical model \cite{Hauser1952}, the cross section of $(p,\alpha)$ reactions scale as

\begin{equation}
   \sigma_{(p,\alpha)} \propto \frac{T_{p0}\,T_{\alpha}}{T_p + T_\gamma + T_\alpha}
       \approx \frac{T_{p0}\,T_{\alpha 0}}{T_{p}} 
       \label{eq:trans}
\end{equation}

The $T_i$ are the transmissions to the open channels, summed over all states $j$ in the residual nuclei: $T_i = \sum_j T_{i.j}$. $T_{p0}$ is the transmission in the entrance channel (with the target $^{59}$Cu in its ground state); $T_\alpha \approx T_{\alpha 0}$ is dominated by the transition to the $^{56}$Ni ground state; excited states are suppressed because of the lower $\alpha$ energy for the population of excited states in the doubly-magic $^{56}$Ni nucleus.

Eq.~(\ref{eq:trans}) indicates that the calculated $(p,\alpha)$ cross section depends on the proton optical model potential (pOMP) (via $T_p$ and $T_{p0}$), the $\alpha$-OMP via $T_{\alpha 0}$, and the level density (ld) in $^{59}$Cu via $T_p$. For further information on the role of the transmissions $T_i$, see \cite{Mohr_EPJA2025_aomp}. The role of the pOMP remains minor because changes in the proton transmission affect $T_{p0}$ and $T_p$ in a similar way, leading to an almost unchanged ratio $T_{p0}/T_p$ in Eq.~(\ref{eq:trans}). 

Thus, to identify the optimal TALYS settings for extrapolating the reaction rate, one has to focus on the $\alpha$-OMP and the ld of $^{59}$Cu. As already noted in \cite{Jaspreet59Cu}, all $\alpha$-OMPs  tend to overestimate the experimental $(p,\alpha)$ cross section, a trend confirmed in the present work.

To identify the model combination best supported by the data, we performed a Bayesian model averaging (BMA) analysis \cite{Hoeting1999} over 96 combinations spanning the $\alpha$-OMP, ld, and $^{59}$Cu maximum discrete-level 
cutoff, with each model scaled to optimally reproduce the data and weighted by its marginal likelihood. The analysis 
favors DEM-3 combined with the Skyrme--Hartree--Fock--Bogolyubov ld 
model~\cite{Goriely_ADNDT2001_HFB} as the statistically preferred configuration and yields an uncertainty factor on the recommended rate ranging from 1.26 to 1.63 over $T_9 = 0.2$--$10$. A  consistent performance of the same Demetriou potential was reported by Gyürky \textit{et al.} for $^{64}$Zn(p,$\alpha$)$^{61}$Cu~\cite{Gy}. 
To further optimize the agreement with the data, a grid search over the geometry parameters of the DEM-3 $\alpha$-OMP was performed. A minor adjustment of the radius (TALYS parameter \texttt{rvadjust}, $+6\%$) and the diffuseness (\texttt{avadjust}, $-6\%$) provides the lowest reduced $\chi^2$ without any post-hoc scaling factor (see Fig.\ref{fig:placeholder_xsec}). Without modifying the ($\alpha$-OMP 5) geometry, reproducing the data requires a 0.86 cross-section scaling factor. Ref.~\cite{LopezSaavedra_Supplemental} details the BMA procedure and $\alpha$-OMP optimization.

The good agreement between the data and the optimized TALYS calculation supports the applicability of the statistical model in the measured energy range. We note that Ref.~\cite{Soltesz} reported a possible anomaly in the $^{60}$Zn ld around 5--6~MeV excitation energy, and could affect the statistical-model rate at $T_9 \lesssim 1$.

Our measurement constrains only the ground-state contribution of $^{59}$Cu to the stellar rate. This reaction rate was obtained using an Exp2Rate~\cite{Rauscher_EXP2RATE_v2.1} calculation that combines the experimental data with the TALYS-predicted cross sections at the optimized $\alpha$-OMP geometry. The stellar rate, which includes contributions from
thermally populated excited states in $^{59}$Cu, was calculated with TALYS at the same optimized geometry.

\begin{figure}[!ht]
    \centering
    \hspace{-2em}
    \includegraphics[width=\linewidth]{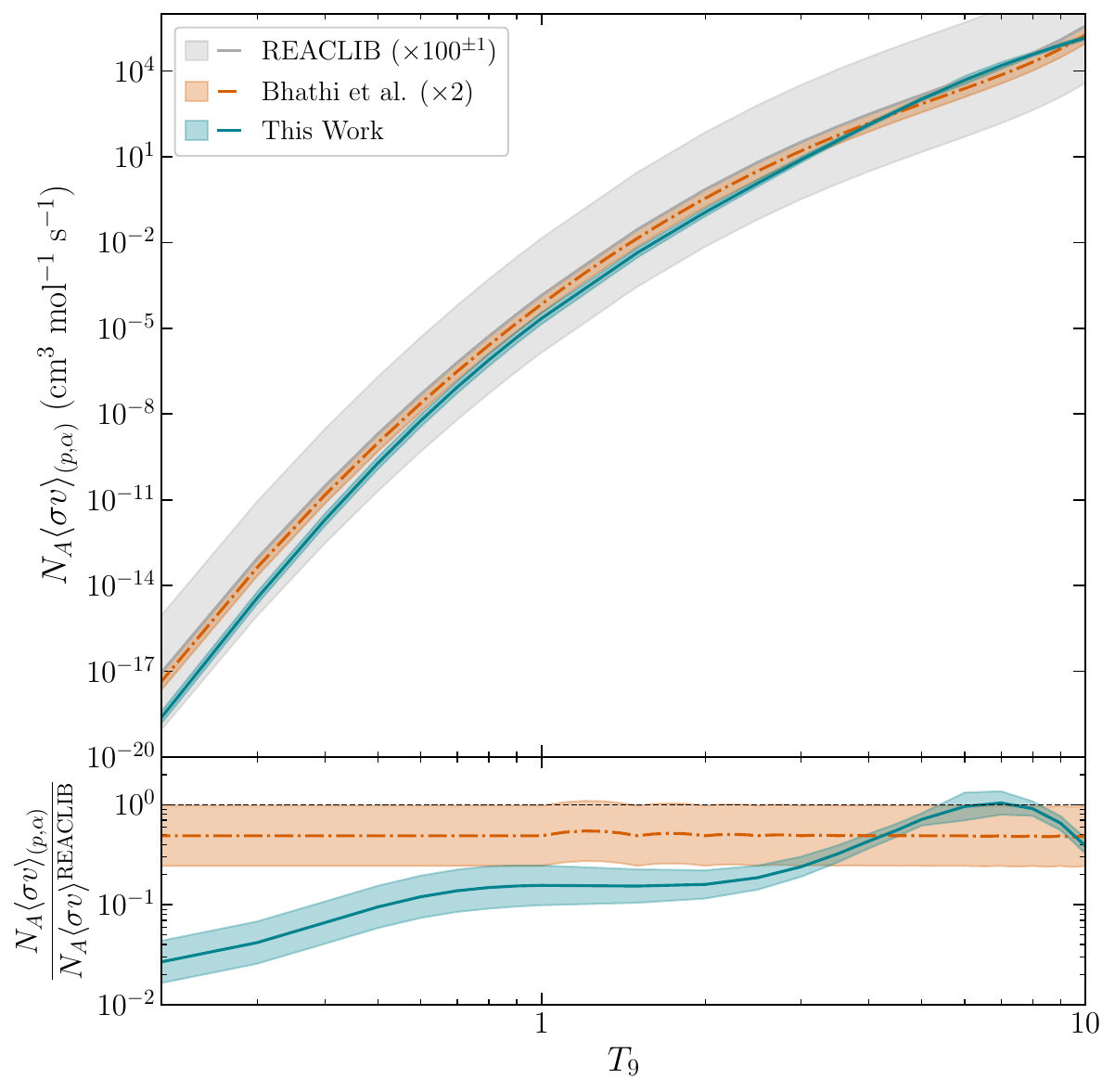}
\caption{(Top) Recommended $^{59}\mathrm{Cu}(p,\alpha)^{56}\mathrm{Ni}$ stellar reaction rate from the present work (teal), compared with Ref.~\cite{Bhathi25} (orange dash-dotted) and REACLIB (gray). Shaded bands indicate the corresponding uncertainties. Bottom: ratio of the rates to REACLIB.}
    \label{fig:rate}
\end{figure}

Fig.~\ref{fig:rate} presents the recommended stellar
$^{59}$Cu$(p,\alpha)^{56}$Ni reaction rate from this work, together with
the rate from Ref.~\cite{Bhathi25} and the REACLIB rate based on
NON\textendash SMOKER calculations~\cite{ths8}, over $T_9 = 0.2$--10.
Because the REACLIB rates are widely used in sensitivity studies
\cite{Parikh_2008,Cyburt_2016,Meisel_2019}, we focus the discussion on a
comparison with REACLIB; the tabulated rate with uncertainty bounds and a
more extended analysis are given in Ref.~\cite{LopezSaavedra_Supplemental}.

The bottom panel of Fig.~2 shows the stellar $^{59}$Cu$(p,\alpha)^{56}$Ni rates from this work and Ref.~\cite{Bhathi25}, relative to REACLIB. At low temperatures, our rate is suppressed by more than an order of magnitude. At XRB and $\nu p$-process temperatures, it remains lower than that of Ref.~\cite{Bhathi25} by a factor of about $2$--$4$, implying a substantially weaker $(p,\alpha)$ channel. This difference reflects the $\alpha$-OMP adopted in Ref.~\cite{Bhathi25}, which is not supported by the present data.

To properly assess the impact of the present results on explosive nucleosynthesis in XRBs and the $\nu$p-process, it is critical to determine the competition between the $(p,\alpha)$ and $(p,\gamma)$ channels. For this comparison, the theoretical $^{59}\mathrm{Cu}(p,\gamma){}^{60}\mathrm{Zn}$ REACLIB rate was used. A recent experimental study by O’Shea \textit{et al.}~\cite{Lotay}, using the $^{59}$Cu$(d,n){}^{60}$Zn reaction to probe proton-unbound resonances in $^{60}$Zn, found a mean $(p,\gamma)$ rate consistent with the REACLIB rate over the relevant temperature range, with uncertainties up to a factor of~5.

The branching ratio $B_{p\alpha/p\gamma}$ provides a measure of the NiCu cycle strength~\cite{Iliadis}. When $B_{p\alpha/p\gamma} > 1$, the cycle is strong and the reaction flow mainly returns to lower masses~\cite{Kim_2022}. When $B_{p\alpha/p\gamma} < 1$, $(p,\gamma)$ breakout dominates, allowing nucleosynthesis to proceed to heavier nuclei.

Fig.~\ref{fig:reactionrateBR} shows $B_{p\alpha/p\gamma}$ calculated with the mean REACLIB $(p,\gamma)$ rate in the denominator and three different $(p,\alpha)$ rates in the numerator: our recommended rate, the REACLIB $(p,\alpha)$ rate varied by a factor of 100 up and down (as used for the sensitivity studies of Refs.~\cite{Cyburt_2016,Meisel_2019}), and the Bhathi \textit{et al.}\cite{Bhathi25} rate with a factor of 2 uncertainty. 
For each rate, the inner and outer bands represent the total uncertainty obtained by adding the $(p,\alpha)$ rate uncertainty in quadrature (assuming the uncertainties are uncorrelated) with assumed $(p,\gamma)$ uncertainty factors of $\times 1$ and $\times 5$~\cite{Lotay}, respectively. This highlights the uncertainty from $(p,\alpha)$ alone and the combined uncertainty from both rates.

\begin{figure}[!htb]
\centering
\includegraphics[width =\linewidth]{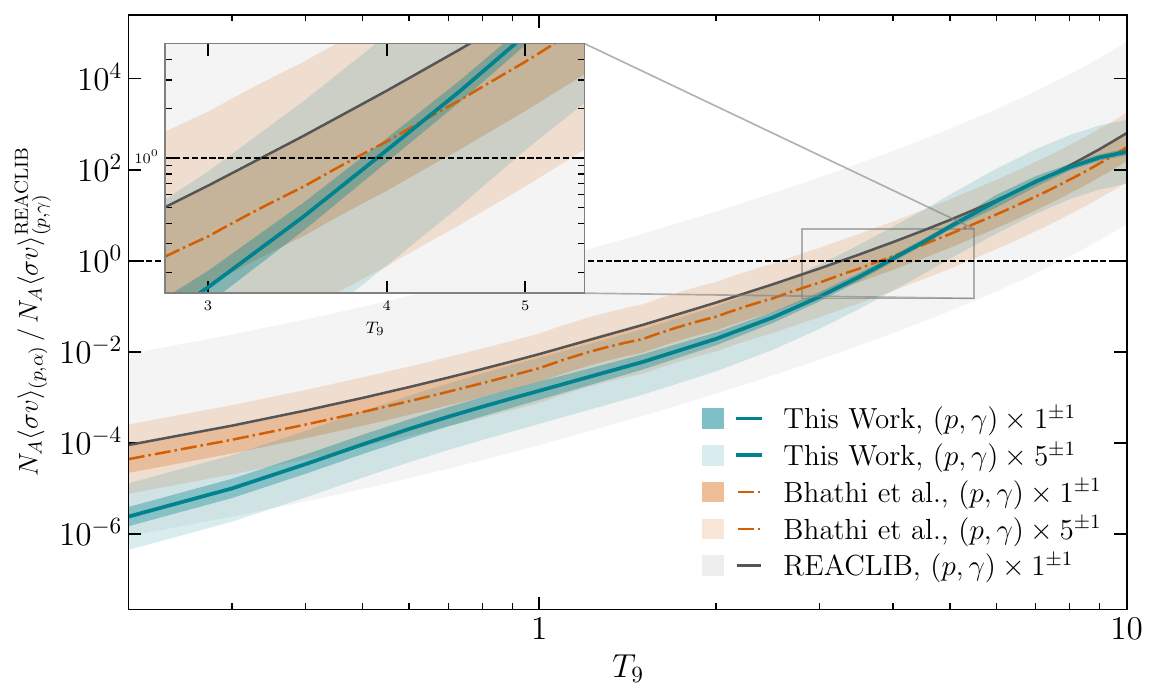}
\label{fig:reactionrateBR}
\caption{Branching ratio $B_{p\alpha/p\gamma}$ vs. $T_9$, using the REACLIB $(p,\gamma)$ rate and the $(p,\alpha)$ rates from this work (teal), REACLIB$\times 100^{\pm 1}$ (gray), and Bhathi \textit{et al.}~\cite{Bhathi25} ($\times 2$) (orange). The dashed line denotes $B=1$; the inset shows the crossover region. Shaded bands show the uncertainties by combining the $(p,\alpha)$ uncertainty in quadrature with assumed $(p,\gamma)$ uncertainty factors of $\times1$ (darker band) and $\times5$ (lighter band).}
\label{fig:reactionrateBR}
\end{figure}

Assuming lognormal rate uncertainties~\cite{Longland2010}, our new $(p,\alpha)$ cross section reduces the $(p,\alpha)$-induced uncertainty on $B_{p\alpha/p\gamma}$ by ${\sim}35$--$60\%$ over $T_9 = 1$--$3$ relative to Ref.~\cite{Bhathi25}. With the $(p,\alpha)$ component now well constrained, the factor-of-5 $(p,\gamma)$ rate uncertainty becomes the dominant contributor to the remaining total error.

At XRB temperatures ($T_9 \lesssim 1$), including a factor-of-5 uncertainty in the $(p,\gamma)$ rate in quadrature constrains the recycling at $T_9=1$ to $0.03$--$0.74\%$. At $T_9=1.5$, we find $<3\%$ recycling, indicating weak NiCu cycle operation under XRB conditions. Ref.~\cite{Bhathi25} reported $<5\%$ at the same temperature; including a factor-of-5 uncertainty in the $(p,\gamma)$ rate increases their result to nearly $10\%$. At $\nu p$-process temperatures ($T_9\approx1$--$3$), the recycling rises to $3$--$45\%$ at $T_9=3$, compared with $5$--$66\%$ from Ref.~\cite{Bhathi25}. Across both regimes, our mean $B_{p\alpha/p\gamma}$ is lower by a factor of $\sim 2$--$4$, indicating a substantially weaker NiCu cycle than previously estimated.

The branching ratio reaches unity at $T_9=3.94^{+0.99}_{-0.85}$ when a factor-of-5 uncertainty in the $(p,\gamma)$ rate is included, compared with $T_9\approx3.2$ predicted by REACLIB. Using instead the $(p,\alpha)$ rate of Ref.~\cite{Bhathi25}, we obtain a crossover temperature of $T_9=3.8^{+1.58}_{-1.22}$. Relative to REACLIB, this higher crossover temperature shifts $\nu p$-process breakout closer to the proto-neutron-star surface, where the stronger neutrino flux enhances the overall efficiency~\cite{Arcones2012NuP}. 
\begin{figure}
    \centering
    \includegraphics[width=\linewidth]{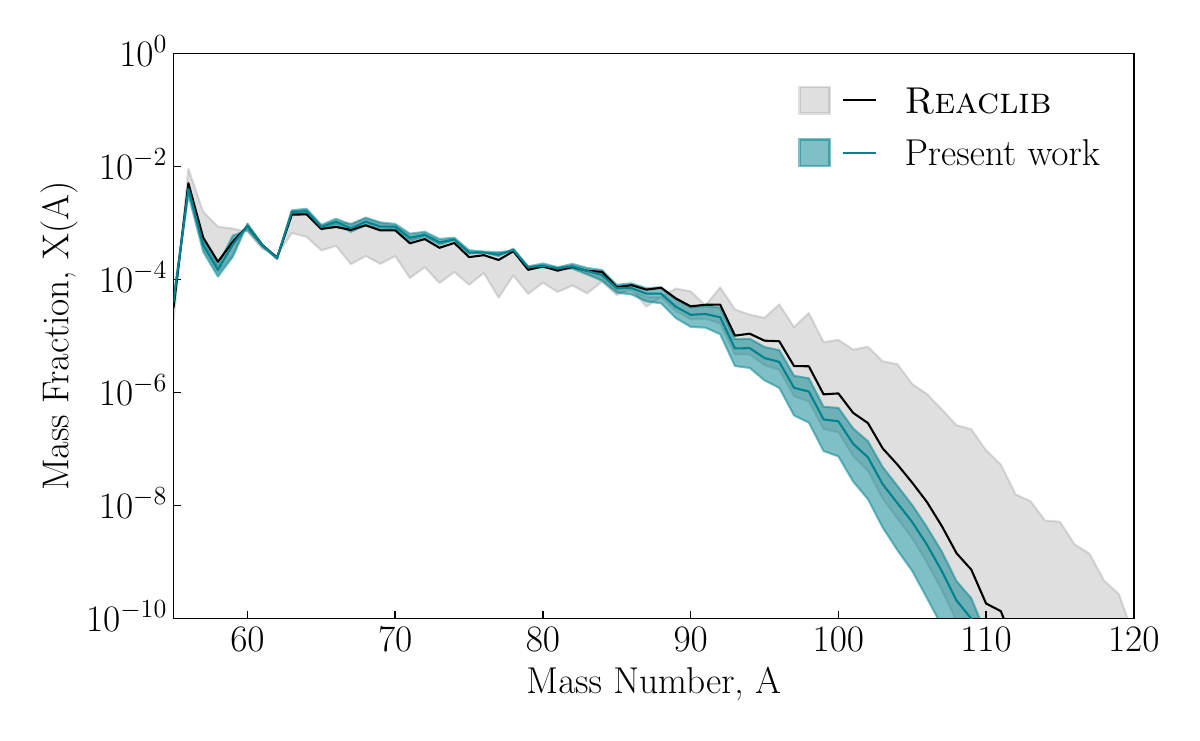}
    \caption{Final $\nu p$-process mass fractions $X(A)$ \cite{PsaltisApJ2024}. The gray band assumes nominal REACLIB rates for $^{59}$Cu($p,\alpha$)$^{56}$Ni ($\times 100^{\pm 1}$) and $^{59}$Cu($p,\gamma$)$^{60}$Zn ($\times 5^{\pm 1}$). The teal band uses our experimental ($p,\alpha$) rate with the same ($p,\gamma$) uncertainty. Solid lines mark central values.}
    \label{fig:vp}
\end{figure}

To assess the astrophysical impact of the newly constrained 
$^{59}$Cu(p,$\alpha$)$^{56}$Ni reaction rate, we performed 
$\nu p$-process nucleosynthesis calculations using a 
representative neutrino-driven wind trajectory from 
Ref.~\cite{PsaltisApJ2024} (Trajectory 40; electron fraction 
$Y_e = 0.56$, entropy $s = 77~k_B$/nucleon, expansion 
timescale $\tau = 19$~ms, and neutron-to-seed ratio 
$\Delta_n = 8.8$). Fig.~\ref{fig:vp} shows the resulting 
final abundance distributions. As a baseline, we adopted 
the nominal REACLIB rates for the competing 
$^{59}$Cu(p,$\alpha$)$^{56}$Ni and $^{59}$Cu(p,$\gamma$)$^{60}$Zn 
reactions assigning uncertainty factors of 100 and 5, 
respectively, to approximate the previously unconstrained 
strength of the NiCu cycle. We then repeated the calculation 
using the present experimental $^{59}$Cu(p,$\alpha$)$^{56}$Ni 
rate while retaining a factor-of-5 uncertainty in the 
competing $(p,\gamma)$ channel. The new measurement 
substantially reduces the uncertainty associated with the 
NiCu cycle and narrows the predicted $\nu p$-process endpoint 
from $A \sim 107$--$119$ to $A \sim 106$--$109$, where we 
define the endpoint as the largest mass number with final 
abundance $X > 10^{-10}$. The reduced $^{59}$Cu(p,$\alpha$)$^{56}$Ni 
strength favors earlier breakout from the NiCu cycle, 
modifying the thermodynamic window over which the 
$\nu p$-process efficiently synthesizes heavy nuclei.

In summary, we report a new direct measurement of the $^{59}$Cu$(p,\alpha)^{56}$Ni excitation function over $E_{\mathrm{cm}}=2.43$--$5.88$ MeV. This provides the first direct experimental input at energies relevant to the $\nu p$ process. Combined with an optimized Hauser--Feshbach analysis that reproduces the data without post-hoc scaling, the measurement reduces the $(p,\alpha)$ rate uncertainty from the factors of $10$--$100$ commonly adopted in sensitivity studies, and from a factor of 2 in Ref.~\cite{Bhathi25} to $1.26$--$1.63$ over $T_9=0.2$--$10$. The resulting stellar rate is systematically lower than previous estimates, demonstrating that the NiCu cycle is substantially weaker than previously inferred. Recycling remains small under XRB conditions, while in the $\nu p$ process the branching crossover shifts upward to $T_9=3.94^{+0.99}_{-0.85}$. The new experimental constraint therefore significantly reduces the uncertainty in the efficiency and endpoint of proton-rich explosive nucleosynthesis. The $^{59}$Cu$(p,\gamma)^{60}$Zn rate now remains the major source of
uncertainty in the NiCu cycle.

\vspace{1 em}
\begin{acknowledgments}
This work is  based upon work supported by the U.S. Department of Energy, Office of Science, Office of Nuclear Physics, under Contract No. DE-AC02-06CH11357 and DE-AC05-00OR22725. It used resources of the Facility for Rare Isotope Beams (FRIB) Operations, which is a DOE Office of Science User Facility under Award Number DE-SC0023633. This work was also supported by the Institute for Basic Science (IBS) funded by the Ministry of Science and ICT, Korea (Grant No. IBS-R031-D1), and by NKFIH (K134197). AP thanks Almudena Arcones for useful discussions and acknowledges the support of the Natural Sciences and Engineering Research Council of Canada (NSERC) under grant SAPIN-2026-00045.
This work was also performed under the auspices of the U.S. Department of Energy by Lawrence Livermore National Laboratory under contract DE-AC52-07NA27344.
\end{acknowledgments}

\FloatBarrier
\bibliography{59Cu_pa}

\end{document}